\documentclass{article}
\usepackage{amssymb}

\input{tcilatex}

\begin{document}

\onecolumn

\begin{titlepage}
\begin{center}
{\LARGE \bf N-body Gravity and the Schroedinger Equation}
\\ \vspace{2cm}
P.S. Farrugia\footnotemark\footnotetext{email: 
pfarruguia@sciborg.uwaterloo.ca},
R.B. Mann\footnotemark\footnotetext{email: 
rbmann@sciborg.uwaterloo.ca},
\\
\vspace{0.5cm} 
Dept. of Physics,
University of Waterloo
Waterloo, ONT N2L 3G1, Canada\\
\vspace{1cm}
T.C. Scott\footnotemark\footnotetext{email: 
scott@pc.rwth-aachen.de}\\
\vspace{0.5cm} 
Institut f\"{u}r Physikalische Chemie, RWTH-Aachen, D-52056 Aachen, Institut f\"{u}r Organische Chemie
Fachbereich Chemie, University of Essen, Germany\\
\vspace{1cm}
PACS numbers: 
13.15.-f, 14.60.Gh, 04.80.+z\\
\vspace{0.5cm}
\today
\vspace{0.5cm}\\
\end{center}

\begin{abstract}

We consider the problem of the motion of $N$ bodies in a self-gravitating system
in two spacetime dimensions. 
We point out that this system can be mapped onto the quantum-mechanical problem
of an N-body generalization of the problem of the H$_{2}^{+}$\ molecular ion in 
one dimension. The canonical gravitational N-body formalism can be extended to include electromagnetic charges. 
We derive a general algorithm for solving this problem, and show how 
it reduces to known results for the 2-body and 3-body systems.
 \end{abstract}
\end{titlepage}\onecolumn

\section{INTRODUCTION}

One of the oldest problems in physics is the N-body problem, which is
concerned with describing the motion of a system of $N$ particles
interacting through specified forces. \ It has found applications across a
very broad range of fields, including astrophysics, condensed matter
physics, plasma physics, nuclear physics and more. \ 

Restricted to the problem of gravitational physics it is particularly
challenging, even in the Newtonian case. \ An exact analytic solution is
only known for the $N=2$ case. For the general relativistic case there is no
exact solution in three spatial dimensions even for $N=2$ (although
approximation techniques exist \cite{THYoG}). \ The main problem here is
taking proper account of the dissipation of energy via gravitational
radiation, and progress here has relied on an eclectic blend of numerical
schemes and approximation techniques.

One of the newest outstanding problems in physics is that of quantum
gravity: finding a consistent and predictive theory of relativistic
gravitation that is fully quantum mechanical and which reduces in the
appropriate (semi-)classical limit to general relativity. \ The most popular
candidates to this end follow the approaches laid out in string theory and
loop quantum gravity. \ However there is currently no fully successful
resolution to the problem.

In both cases considerable insight has been attained by reducing the number
of spacetime dimensions. \ Nonrelativistic self-gravitating systems (OGS) of
N particles in one spatial dimension have been very important in the fields
of astrophysics and cosmology for over 30 years \cite{Rybicki}, and
2-dimensional quantum gravity has been studied for over 25 years \cite{JT}.
\ In the latter case a broad variety of possible theories presents itself. \
This is because the Einstein-Hilbert action is a topological invariant in
two spacetime dimensions, and therefore has trivial field equations. \ It is
therefore necessary to modify two-dimensional relativistic gravity in some
way, and the most common procedure for doing so is to incorporate a scalar
(dilaton) field. \ This has the effect of yielding non-trivial equations of
motion, albeit at the price of indissoluably coupling the dilaton to gravity
and all of the other matter fields in the system. Though this general
procedure has antecedants from string theory, it bears a limited resemblance
to (3+1)-dimensional general relativity insofar as the latter has no such
dilatonic couplings (though we note that the s-wave sector of ($D+1$%
)-dimensional general relativity is reproduced by specific dilaton gravity
models \cite{Grumiller}). Furthermore the nonrelativistic ($c\rightarrow
\infty $) limits of such theories in general bear little resemblance to
(1+1)-dimensional non-relativistic gravity \cite{jchan}.

\bigskip

However there is a particular way of incorporating the dilaton that
successfully addresses both concerns. \ The approach involves choosing the
coupling so that the field equations couple the stress-energy $T_{\mu
\upsilon }$ of (non-dilatonic) matter to curvature in a manner analogous to
that of Einstein gravity in (3+1) dimensions. \ Since the only measure of
curvature in two spacetime dimensions is the Ricci scalar $R$, it is only
the trace of the stress-energy that can act as the source for curvature,
suggesting the equation 
\begin{equation}
R=\kappa T_{\mu }^{\text{ \ \ }\mu }
\end{equation}%
Consequently, as in (3+1) dimensions, the evolution of space-time curvature
is governed by the matter distribution, which in turn is governed by the
dynamics of space-time \cite{r3}. \ Referred to as $R=T$\ theory, it is a
particular member of a class of dilation gravity theories on a line (or a
circle, pending the choice of topology for an initial data set). \ Long
regarded as a model quantum theory of gravity when $T_{\mu }^{\text{ \ \ }%
\mu }$ is constant \cite{JT}, what makes it particularly interesting is that
its non-relativistic limit is that of the Newtonian N-body system when the
stress-energy is that of N point particles minimally coupled to gravity.
Indeed, it can be regarded as the two-dimensional limit to general
relativity \cite{Ross2d}.

The $R=T$ theory therefore forms an ideal theoretical laboratory not only
for quantum gravity, but also for studying both the OGS and its relativistic
counterparts. In the latter case there are physical systems with dynamics
closely approximated by the one dimensional system. For example\ very
long-lived core-halo configurations are known to exist in the OGS phase
space \cite{yawn}. These are reminiscent of structures observed in globular
clusters and model a dense massive core in near-equilibrium that is
surrounded by a halo of high kinetic energy stars that interact only weakly
with the core. \ Other higher-dimensional referents include flat parallel
domain walls moving in a direction perpendicular to their surfaces and the
dynamics of stars in a direction orthogonal to the plane of a highly
flattened galaxy. \ The relativistic OGS\ (ROGS) has yielded additional
insight as to how these properties are modified by general-relativistic
effects. \ The two-body problem has been solved exactly in a variety of
physical settings. \ The 3-body problem has been solved numerically in both
the equal \cite{3bdshort} and unequal \cite{Justin} mass cases. \ Remarkably
such a system shows no evidence of increased chaotic behaviour relative to
its non-relativistic counterpart, despite the high degree of non-linearity
in the system \cite{burnell} .

\bigskip

The purpose of this paper is to explore the connection between this theory
and its potential quantum-mechanical counterpart. \ We note in particular
that the constraint equation of the gravity theory is identical to the
stationary Schroedinger equation for a N-body linear atom. \ We show how to
solve these constraint equations to obtain an implicit expression (called
the determining equation) for the Hamiltonian in the N-body case,
generalizing an approach developed in the 3-body system \cite{3bdshort}. In
the 2-body case the determining equation can be explicitly solved since it
reduces to the defining equation for the Lambert-W function. This yields an
expression for the Hamiltonian as a function of the separation of the
particles and its conjugate momentum. In the 3-body case the determining
equation naturally suggests a generalization of the Lambert W-function,
whose properties have been outlined elsewhere \cite{MMS}. The Lambert W
function appears in a number of fundamental problems, such as the nodal
structures of wave functions of atoms\cite{nodal}.

The outline of our paper is as follows. We begin by reviewing the
formulation of the $N$-body problem in (1+1)-dimensional relativistic
gravity. The problem reduces to that of solving a single constraint
equation. \ We then discuss how this equation can be mapped onto the
Schroedinger equation for an $N$-body generalization of the H$_{2}^{+}$\
molecular ion. \ We then provide a procedure for solving the constaint
equation for an arbitrary number of bodies, and show how it reduces to
previously known results for $N=2$\ and $N=3$. We summarize our work in a
concluding section.

\section{Basics\textit{\qquad }}

The action integral for the gravitational field coupled with N point
particles in two spacetime dimensions is \cite{3bdshort,OR,2bd}%
\begin{eqnarray}
I &=&\int d^{2}x\left[ \frac{1}{2\kappa }\sqrt{-g}\left\{ \Psi R+\frac{1}{2}%
g^{\mu \nu }\bigtriangledown _{\mu }\Psi \bigtriangledown _{\nu }\Psi
+\Lambda \right\} \right.  \label{eqn-act1} \\
&&\left. -\sum\limits_{a=1}^{N}m_{a}\int d\tau _{a}\left\{ -g_{\mu \nu }(x)%
\frac{dz_{a}^{\mu }}{d\tau _{a}}\frac{dz_{a}^{\nu }}{d\tau _{a}}\right\} ^{%
\frac{1}{2}}\delta ^{2}(x-z_{a}(\tau _{a}))\right]  \nonumber
\end{eqnarray}%
where\ $R$ is the Ricci scalar, $g_{\mu \nu }$ and $g$ are the metric and
its determinant, $\tau _{a}$ is the proper time of the $a$-th particle, $%
\kappa =8\pi G/c^{4}$ is the gravitational coupling, and $\Psi $ is a scalar
field called the dilaton. \ This action describes a generally covariant
self-gravitating system (without collision terms, so that the bodies pass
through each other), in which the scalar curvature is sourced by the point
particles and the cosmological constant $\Lambda $.

The action (\ref{eqn-act1}) neglects the inclusion of boundary terms, and so
is only correct if the spacetime is compact without boundaries. When
boundaries are present one must include both the analogue of the
Gibbons-Hawking-York term with extrinsic curvature \cite{KCK}, but also a
Hamilton-Jacobi counterterm, which recently has been calculated from first
principles \cite{GrumMc}. Since we are interested in the methods required
for obtaining a general solution to the field equations for the N-body
problem, these terms are not relevant for our purposes. \ For a more
complete discussion of boundary terms for the N-body problem see ref \cite%
{Geoff}.

The field equations are obtained by varying the action with respect to the
metric, dilation field, and particle coordinates. After some manipulation
this gives

\begin{equation}
R-\Lambda =\kappa T_{\mu }^{P\mu }  \label{eqn-rt}
\end{equation}

\begin{equation}
\frac{d}{d\tau _{a}}\left\{ \frac{dz_{a}^{\nu }}{d\tau _{a}}\right\} +\Gamma
_{\alpha \beta }^{\nu }\left( z_{a}\right) \frac{dz_{a}^{\alpha }}{d\tau _{a}%
}\frac{dz_{a}^{\beta }}{d\tau _{a}}=0  \label{eqn-z}
\end{equation}

\begin{equation}
\frac{1}{2}\bigtriangledown _{\mu }\Psi \bigtriangledown _{\nu }\Psi -g_{\mu
\nu }(\frac{1}{4}\bigtriangledown ^{\lambda }\Psi \bigtriangledown _{\lambda
}\Psi -\bigtriangledown ^{2}\Psi )-\bigtriangledown _{\mu }\bigtriangledown
_{\nu }\Psi =\kappa T_{\mu \nu }^{P}+\frac{\Lambda }{2}g_{\mu \nu }
\label{eqn-psi}
\end{equation}%
where the stress-energy due to the point masses is

\begin{equation}
T_{\mu \nu }^{P}=\sum_{a=1}^{N}m_{a}\int d\tau _{a}\frac{1}{\sqrt{-g}}g_{\mu
\sigma }g_{\nu \rho }\frac{dz_{a}^{\sigma }}{d\tau _{a}}\frac{dz_{a}^{\rho }%
}{d\tau _{a}}\delta ^{2}\left( x-z_{a}\left( \tau _{a}\right) \right)
\label{eqn-stressenergy}
\end{equation}%
and is conserved. \ Note that (\ref{eqn-rt},\ref{eqn-z}) are a closed system
of $N+1$ equations, which can be solved for the N degrees of freedom of the
point masses and the single metric degree of freedom. \ Consistent with the
conservation of $T_{\mu \nu }$, the left-hand side of (\ref{eqn-stressenergy}%
) is divergenceless, yielding only one independent equation to determine the
single degree of freedom of the dilaton, whose evolution is thus governed by
the evolution of the point masses via (\ref{eqn-psi}). \ 

\qquad Making use of the decomposition 
\begin{equation}
\sqrt{-g}R=-2\partial _{0}(\sqrt{\gamma }K)+2\partial _{1}(\sqrt{\gamma }%
N^{1}K-\gamma ^{-1}\partial _{1}N_{0})  \label{decompK}
\end{equation}
where the extrinsic curvature is%
\begin{equation}
K=(2N_{0}\gamma )^{-1}(2\partial _{1}N_{1}-\gamma ^{-1}N_{1}\partial
_{1}\gamma -\partial _{0}\gamma )  \label{extK}
\end{equation}
we can rewrite the action in the canonical form

\begin{equation}
I=\int dx^{2}\left\{ \sum_{a}p_{a}\dot{z}_{a}\delta \left( x-z_{a}\left(
x^{0}\right) \right) +\pi \dot{\gamma}+\Pi \dot{\Psi}+N_{0}R^{0}+N_{1}R^{1}%
\right\}  \label{eqn-act2}
\end{equation}%
where $\gamma =g_{11}$, $N_{0}=(-g^{00})^{-\frac{1}{2}}$, $N_{1}=g_{10}$,
and $\pi $ and $\Pi $ are conjugate momenta to $\gamma $ and $\Psi $
respectively. \ The quantities $N_{0}$ and $N_{1}$ are Lagrange multipliers
that enforce the constraints $R^{0}=0=R^{1}$, where%
\begin{equation}
R^{0}=-\kappa \sqrt{\gamma }\gamma \pi ^{2}+2\kappa \sqrt{\gamma }\pi \Pi +%
\frac{\left( \Psi ^{\prime }\right) ^{2}}{4\kappa \sqrt{\gamma }}-\left( 
\frac{\Psi ^{\prime }}{\kappa \sqrt{\gamma }}\right) ^{\prime }+\frac{%
\Lambda }{2\kappa }\sqrt{\gamma }-\sum_{a}\sqrt{\frac{p_{a}^{2}}{\gamma }%
+m_{a}^{2}}\delta \left( x-z_{a}\left( x^{0}\right) \right)  \label{eqn-R0}
\end{equation}%
\begin{equation}
R^{1}=\frac{\gamma ^{\prime }}{\gamma }\pi -\frac{1}{\gamma }\Pi \Psi
^{\prime }+2\pi ^{\prime }+\sum_{a}\frac{p_{a}}{\gamma }\delta \left(
x-z_{a}\left( x^{0}\right) \right)  \label{eqn-R1}
\end{equation}%
with the symbols ( $^{\cdot }$\ ) and ( $^{\prime }$ ) denoting $\partial
_{0}$ and $\partial _{1}$, respectively.

The dynamical and gauge degrees of freedom\ can be identified by writing \
the generator arising from the variation of the action at the boundaries in
terms of $\left( \Psi ^{\prime }/\sqrt{\gamma }\right) ^{\prime }$ and $\pi
^{\prime }$. The coordinate conditions can be chosen to be $\gamma =1$ and $%
\Pi =0$; upon elimination of the constraints, the action becomes%
\begin{equation}
I=\int d^{2}x\left\{ \sum_{a}p_{a}\dot{z}_{a}\delta \left( x-z_{a}\right) -%
\mathcal{H}\right\}   \label{eqn-act3}
\end{equation}%
where 
\begin{equation}
H=\int dx\mathcal{H}=-\frac{1}{\kappa }\int dx\Delta \Psi   \label{Ham1}
\end{equation}%
is the reduced Hamiltonian, with $\Delta \equiv \partial ^{2}/\partial x^{2}$%
. The field $\Psi =\Psi \left( x,z_{a},p_{a}\right) $ is understood to be
determined from the constraint equations which are now

\begin{equation}
\Delta \Psi -\frac{\left( \Psi ^{\prime }\right) ^{2}}{4}+\kappa ^{2}\pi
^{2}-\frac{\Lambda }{2}+\kappa \sum_{a}\sqrt{p_{a}^{2}+m_{a}^{2}}\delta
\left( x-z_{a}\right) =0  \label{cst1}
\end{equation}%
\begin{equation}
2\Delta \chi +\sum_{a}p_{a}\delta \left( x-z_{a}\right) =0  \label{cst2}
\end{equation}%
where $\pi =\chi ^{\prime }$. \ The consistency of this canonical reduction
can be demonstrated \cite{OR}\ by showing that the canonical equations of
motion derived from the reduced Hamiltonian (\ref{Ham1}) are identical with
the canonical field equations.

The procedure for obtaining a solution to the system is as follows.
Equations (\ref{cst1},\ref{cst2}) first must be solved to obtain $\Psi $\ as
a function of the phase-space variables $\left( z_{a},p_{a}\right) $,
subject to the boundary conditions that the Hamiltonian (\ref{Ham1}) is
finite for large $\left| x\right| $. Insertion of this result into (\ref%
{Ham1}) then yields the Hamiltonian as function of the physical degrees of
freedom $\left( z_{a},p_{a}\right) $\ of the system. \ Hamilton's equations
then can be employed to solve for the evolution of the system.\ Using these
results, the remaining components of the metric can be obtained by solving
the field equations that follow from (\ref{eqn-act2}) \cite{2bd}. \ 

Note that in $(1 + 1)$ dimensions the dimensionless potential$Gmr/c^{2}$\
between a pair of particles separated by a distance $r$\ becomes infinite at
spatial infinity, in contrast to (3+1) dimensions where the corresponding
quantity $Gm/rc^{2}$\ vanishes at spatial infinity. \ Consequently the
finiteness of the Hamiltonian in (1+1) dimensions must be ensured by an
appropriate choice of boundary condition on the variables $\Psi $\ and $\chi 
$. \ This condition has been shown to be \cite{OR,2bd} $\Psi ^{2}-4\kappa
^{2}\chi ^{2}=0$\ for large large $\left| x\right| $, and guarantees both
that the surface terms that arise in the action vanish and that the
Hamiltonian is finite for the $N$-body system.

\section{Comparison with the Schroedinger Equation}

Eq. (\ref{cst1}) is the same as the stationary Schroedinger equation for a
single-electron atom with $N$ field sources. Consider first the situation $%
z_{n}<z_{n-1}<\ldots <z_{1}$. This divides the region into $n+1$ regions
defined by 
\[
\begin{array}{l}
\text{Region \ \ \ \ \ \ \ \ \ \ }0:\quad z_{1}<x \\ 
\text{Region \ \ \ \ \ \ \ \ \ \ \ }i:\quad z_{i+1}<x<z_{i} \\ 
\text{Region \ \ \ \ \ \ \ \ \ \ }n:\quad x<z_{n}%
\end{array}%
\]%
Setting $E_{pi}=\sqrt{p_{i}^{2}+m_{i}^{2}}$ and $\Psi =-4\ln |\phi |$ allows
us to rewrite (\ref{cst1}) as%
\begin{equation}
\bigtriangledown ^{2}\phi -\frac{1}{4}\left[ \kappa ^{2}(\chi ^{\prime
})^{2}-\frac{\Lambda }{2}+\kappa E_{pi}\delta (x-z_{i})\right] \phi (x)=0
\label{cstphi}
\end{equation}%
where the summation convention applies to the repeated Latin index.

Eqn. (\ref{cst2}) has the general solution 
\begin{equation}
\chi =-\frac{1}{4}p_{i}|x-z_{i}|-\epsilon Xx+\epsilon C_{\chi }.
\label{chieqn}
\end{equation}%
where $X$ and $C_{\chi }$ are integration constants. The factor of $\epsilon 
$ appears so that behaviour of $\chi $ is odd under time reversal; when $%
t\rightarrow -t$ , $\epsilon \rightarrow -\epsilon $. In each region $r$,
the function $\chi ^{\prime }$ is spatially constant 
\begin{equation}
\chi ^{\prime }=-\epsilon X-\frac{1}{4}p_{i}s_{pi,r}  \label{chiprime}
\end{equation}%
where we have defined 
\[
s_{pi,r}=\left\{ 
\begin{array}{rl}
\;1 & \text{if}\;i>r \\ 
-1 & \text{if}\;i\leq r%
\end{array}%
\right. 
\]

Insertion of (\ref{chiprime}) into eqn. (\ref{cstphi}) yields 
\begin{equation}
\bigtriangledown ^{2}\phi -\frac{1}{4}\left[ \kappa ^{2}(\epsilon X+\frac{1}{%
4}p_{i}s_{pi,r})^{2}-\frac{\Lambda }{2}+\kappa E_{pi}\delta (x-z_{i})\right]
\phi (x)=0
\end{equation}%
or alternatively%
\begin{equation}
-\bigtriangledown ^{2}\phi +\frac{1}{4}\left[ \kappa E_{pi}\delta (x-z_{i})%
\right] \phi (x)=\mathcal{E}\left( E_{pi}\right) \phi (x)  \label{schgr}
\end{equation}%
where 
\begin{equation}
\mathcal{E}\left( E_{pi}\right) =\frac{\Lambda }{8}-\frac{\kappa ^{2}}{4}%
(X\left( E_{pi}\right) +\frac{1}{4}\epsilon p_{i}s_{pi,r})^{2}
\label{eigenval}
\end{equation}

Equation (\ref{schgr}) is formally the same as the stationary Schroedinger
equation for a particle of mass $m$ and energy $\frac{\hslash ^{2}}{2m}%
\mathcal{E}$ interacting with a linear molecule whose field sources have
charges $\kappa E_{pi}$. \ This is an N-body generalization of the problem
of the H$_{2}^{+}$\ molecular ion in one dimension\cite%
{Frost,Whitton,ScottBabb}. \ We note that there exists a D-dimensional
version of H$_{2}^{+}$\ , which can be obtained via a scheme called
`dimensional scaling' \cite{Frantz,Lopez,ScottFrecon}. The 2-body case of
the one-dimensional limit of H$_{2}^{+}$\ is solvable in terms of a
generalized Lambert W function \cite{2bd}.

It is important to note that this analysis can be extended to include an $%
(1+1)$-dimensional electromagnetic field between the interacting (charged)
particles\cite{EM}. In this case, eq.~(\ref{cstphi}) is generalized to %
\cite[(46)]{EM}: 
\begin{equation}
\bigtriangledown ^{2}\phi -\frac{1}{4}\left[ \kappa ^{2}(\chi ^{\prime
})^{2}+\frac{\kappa }{2}V-\frac{\Lambda _{e}}{2}+\kappa E_{pi}\delta
(x-z_{i})\right] \phi (x)=0  \label{cstphi-em}
\end{equation}%
where the potential $V(x)\equiv E^{2}-C$, $E$ is the electric field and $%
\Lambda _{e}$ is the effective cosmological constant $\Lambda _{e}\equiv
\Lambda -\kappa C$ where $C\equiv =\textstyle\frac{1}{4}(\sum_{a}e_{a})^{2}$
where $e_{a}$ is charge associated with the $a^{th}$ particle as defined in %
\cite[(40)]{EM}. Thus, in the presence of an electromagnetic field, eqs.~(%
\ref{schgr}) and (\ref{eigenval}) are generalized to 
\begin{equation}
-\bigtriangledown ^{2}\phi +\frac{1}{4}\left[ \frac{\kappa }{2}V+\kappa
E_{pi}\delta (x-z_{i})\right] \phi (x)=\mathcal{E}\left( E_{pi}\right) \phi
(x)  \label{schgr-em}
\end{equation}%
where 
\begin{equation}
\mathcal{E}\left( E_{pi}\right) =\frac{\Lambda _{e}}{8}-\frac{\kappa ^{2}}{4}%
(X\left( E_{pi}\right) +\frac{1}{4}\epsilon p_{i}s_{pi,r})^{2}
\label{eigenval-em}
\end{equation}%
Note that when the gravitational coupling $\kappa $ is weak relative to the
electromagnetic couplings, we recover the conventional Schr\"{o}dinger
equation for electromagnetically charged particles.

\section{Solving the Constraint Equations}

In this section we outline the general procedure for solving the constraint
equations (\ref{cst1}) and (\ref{cst2}) in the $N$-body case.

Eqn. (\ref{cstphi}) when $x\neq z_{a}$ 
\begin{equation}
\bigtriangledown ^{2}\phi +\frac{1}{4}\phi \biggl[\kappa ^{2}(\chi ^{\prime
})^{2}-\frac{\Lambda }{2}\biggr]=\bigtriangledown ^{2}\phi +\frac{1}{4}\phi %
\biggl[\kappa ^{2}(-\epsilon X-\frac{1}{4}p_{i}s_{pi,r})^{2}-\frac{\Lambda }{%
2}\biggr]=0
\end{equation}%
can be solved in each region, yielding, 
\begin{equation}
\phi _{r}=A_{r}e^{\frac{1}{2}K_{r}x}+B_{r}e^{-\frac{1}{2}K_{r}x}
\label{phirsol}
\end{equation}%
where%
\begin{equation}
K_{r}=\sqrt{\kappa ^{2}(X+\frac{1}{4}\epsilon p_{i}s_{pi,r})^{2}-\frac{%
\Lambda }{2}}  \label{KEqn}
\end{equation}%
The solutions (\ref{phirsol}) can be regarded as `free particle' solutions.
We procede in the same fashion as for a quantum mechanics problem with field
sources that are Dirac delta functions. Outside these sources, the particles
are `free' since there is no potential.

This set of equations must satisfy the matching conditions at the locations $%
x=z_{{i}}$. These are%
\begin{eqnarray}
\phi _{r-1}(z_{r}) &=&\phi _{r}(z_{r})=\phi (z_{r})  \label{ConditionPhi1} \\
\phi _{r-1}^{\prime }(z_{r})-\phi _{r}^{\prime }(z_{r}) &=&\frac{1}{4}\kappa
E_{pr}\phi (z_{r})  \label{ConditionPhi2}
\end{eqnarray}%
\\*[0pt]
From conditions (\ref{ConditionPhi1}) and (\ref{ConditionPhi2})%
\begin{eqnarray}
A_{r-1}e^{\frac{1}{2}K_{r-1}z_{r}}+B_{r-1}e^{-\frac{1}{2}K_{r-1}z_{r}}
&=&A_{r}e^{\frac{1}{2}K_{r}z_{r}}+B_{r}e^{-\frac{1}{2}K_{r}z_{r}}
\label{ConditionPhi3} \\
A_{r-1}e^{\frac{1}{2}K_{r-1}z_{r}}-B_{r-1}e^{-\frac{1}{2}K_{r-1}z_{r}} &=&%
\frac{\kappa E_{pr}+2K_{r}}{2K_{r-1}}A_{r}e^{\frac{1}{2}K_{r}z_{r}}+\frac{%
\kappa E_{pr}-2K_{r}}{2K_{r-1}}B_{r}e^{-\frac{1}{2}K_{r}z_{r}}  \nonumber \\
&&  \label{ConditionPhi-4}
\end{eqnarray}%
Adding eqns (\ref{ConditionPhi3}) and (\ref{ConditionPhi-4}) gives 
\begin{equation}
A_{r-1}e^{\frac{1}{2}K_{r-1}z_{r}}=\frac{\kappa E_{pr}+2(K_{r}+K_{r-1})}{%
4K_{r-1}}A_{r}e^{\frac{1}{2}K_{r}z_{r}}+\frac{\kappa E_{pr}-2(K_{r}-K_{r-1})%
}{4K_{r-1}}B_{r}e^{-\frac{1}{2}K_{r}z_{r}}  \label{ConditionPhi5}
\end{equation}%
whereas 
\begin{equation}
B_{r-1}e^{-\frac{1}{2}K_{r-1}z_{r}}=-\frac{\kappa E_{pr}+2(K_{r}-K_{r-1})}{%
4K_{r-1}}A_{r}e^{\frac{1}{2}K_{r}z_{r}}-\frac{\kappa E_{pr}-2(K_{r}+K_{r-1})%
}{4K_{r-1}}B_{r}e^{-\frac{1}{2}K_{r}z_{r}}  \label{ConditionPhi6}
\end{equation}%
is obtained upon subtracting them.\\*[0pt]

The boundary conditions that ensure the Hamiltonian is finite in the regions 
$0$ and $n$ is, 
\begin{equation}
\Psi ^{2}-4\kappa ^{2}\chi ^{2}+2\Lambda x^{2}=C_{0/n}x
\label{BoundaryInfCond1}
\end{equation}%
where $C_{0}$ and $C_{n}$ are constants that have to be determined. Eqn. (%
\ref{BoundaryInfCond1}) immediately gives%
\begin{equation}
A_{n}=B_{0}=0  \label{BoundaryInfCond2}
\end{equation}

Using Eqn. (\ref{BoundaryInfCond2}) in eqns (\ref{ConditionPhi5}) and (\ref%
{ConditionPhi6}) with $r=n$ then implies%
\begin{eqnarray}
A_{n-1} &=&\frac{\kappa E_{pn}-2(K_{n}-K_{n-1})}{4K_{n-1}}B_{n}e^{-\frac{1}{2%
}(K_{n}+K_{n-1})z_{n}}\equiv \frac{M_{An}}{4K_{n-1}}B_{n}
\label{ConditionPhi7} \\
B_{n-1} &=&-\frac{\kappa E_{pn}-2(K_{n}+K_{n-1})}{4K_{n-1}}B_{n}e^{-\frac{1}{%
2}(K_{n}-K_{n-1})z_{n}}\equiv \frac{M_{Bn}}{4K_{n-1}}B_{n}
\label{ConditionPhi8}
\end{eqnarray}%
Defining%
\begin{eqnarray}
L_{AAr} &=&\kappa E_{pr}+2(K_{r}+K_{r-1})\qquad L_{ABr}=\kappa
E_{pr}-2(K_{r}-K_{r-1}) \\
L_{BAr} &=&\kappa E_{pr}+2(K_{r}-K_{r-1})\qquad L_{BBr}=\kappa
E_{pr}-2(K_{r}+K_{r-1})
\end{eqnarray}%
\begin{eqnarray}
e_{AAr} &=&e^{\frac{1}{2}(K_{r}-K_{r-1})z_{r}}\qquad e_{ABr}=e^{-\frac{1}{2}%
(K_{r}+K_{r-1})z_{r}} \\
e_{BAr} &=&e^{\frac{1}{2}(K_{r}+K_{r-1})z_{r}}\qquad e_{BBr}=e^{-\frac{1}{2}%
(K_{r}-K_{r-1})z_{r}}
\end{eqnarray}%
allows us to write eqns. (\ref{ConditionPhi5}) and (\ref{ConditionPhi6}) as%
\begin{eqnarray}
A_{r-1} &=&\frac{1}{4K_{r-1}}(L_{AAr}A_{r}e_{AAr}+L_{ABr}B_{r}e_{ABr}) \\
B_{r-1} &=&\frac{1}{4K_{r-1}}(-L_{BAr}A_{r}e_{BAr}-L_{BBr}B_{r}e_{BBr})
\end{eqnarray}%
For $r=n-1,$ eqn.(\ref{ConditionPhi7}, \ref{ConditionPhi8}) give%
\begin{eqnarray}
A_{n-2} &=&\frac{1}{4^{2}K_{n-1}K_{n-2}}%
(L_{AA,n-1}M_{An}e_{AA,n-1}+L_{AB,n-1}M_{Bn}e_{AB,n-1})B_{n}  \nonumber \\
&=&\frac{M_{A,n-1}}{4^{2}K_{n-1}K_{n-2}}B_{n} \\
B_{n-2} &=&\frac{1}{4^{2}K_{n-1}K_{n-2}}%
(-L_{BA,n-1}M_{An}e_{BA,n-1}-L_{BB,n-1}M_{Bn}e_{BB,n-1})B_{n}  \nonumber \\
&=&\frac{M_{B,n-1}}{4^{2}K_{n-1}K_{n-2}}B_{n}
\end{eqnarray}%
Repeating this for $r=n-2$ yields%
\begin{eqnarray}
A_{n-3} &=&\frac{1}{4^{3}K_{n-1}K_{n-2}K_{n-3}}%
(L_{AA,n-2}M_{A,n-1}e_{AA,n-2}+L_{AB,n-2}M_{B,n-1}e_{AB,n-2})B_{n}  \nonumber
\\
&=&\frac{M_{A,n-2}}{4^{3}K_{n-1}K_{n-2}K_{n-3}}B_{n} \\
B_{n-3} &=&\frac{1}{4^{3}K_{n-1}K_{n-2}K_{n-3}}%
(-L_{BA,n-2}M_{A,n-1}e_{BA,n-2}-L_{BB,n-2}M_{B,n-1}e_{BB,n-2})B_{n} 
\nonumber \\
&=&\frac{M_{B,n-2}}{4^{3}K_{n-1}K_{n-2}K_{n-3}}B_{n}
\end{eqnarray}

In this way a pattern clearly emerges, giving the general form for $A_{n-r}$
and $B_{n-r}$ as, 
\begin{eqnarray}
A_{n-r} &=&\frac{M_{A,n-r+1}}{4^{r}\prod_{j=1}^{r}K_{n-j}}B_{n} \\
B_{n-r} &=&\frac{M_{B,n-r+1}}{4^{r}\prod_{j=1}^{r}K_{n-j}}B_{n},
\end{eqnarray}%
where $M_{A,n-r+1}$ and $M_{B,n-r+1}$ are defined by the recurrence
relations, 
\begin{eqnarray}
M_{A,n-r} &=&L_{AA,n-r}M_{A,n-r+1}e_{AA,n-r}+L_{AB,n-r}M_{B,n-r+1}e_{AB,n-r}
\label{RecRelationMB} \\
M_{B,n-r} &=&-L_{BA,n-r}M_{A,n-r+1}e_{BA,n-r}-L_{BB,n-r}M_{B,n-r+1}e_{BB,n-r}
\label{RecRelationMA} \\
&&  \nonumber
\end{eqnarray}%
with 
\[
M_{A,n+1}=0\qquad \qquad M_{B,n+1}=1
\]%
Evaluating the recurrence relations (\ref{RecRelationMA}) and (\ref%
{RecRelationMB}) for the case $r=n$, and resorting to eqns. (\ref%
{BoundaryInfCond2}) furnishes the relation%
\[
0=\frac{M_{B,0}}{4^{n+1}\prod_{j=1}^{n+1}K_{n-j}}B_{n}\qquad \Longrightarrow
\qquad M_{B,1}=0
\]%
since $B_{n}$ cannot be equal to zero.

Furthermore%
\[
H=-\frac{1}{\kappa }\int dx\bigtriangledown ^{2}\Psi =-\frac{1}{\kappa }%
\biggr[\Psi ^{\prime }\biggl]_{\infty }^{\infty }=-\frac{4}{\kappa }\biggr[%
\frac{\phi ^{\prime }}{\phi }^{\prime }\biggl]_{\infty }^{\infty }=-\frac{4}{%
\kappa }\frac{1}{2}(-K_{0}-K_{n}) 
\]%
\begin{equation}
H=\frac{2}{\kappa }(K_{0}+K_{n})=\frac{4}{\kappa }\sqrt{\kappa ^{2}X^{2}-%
\frac{\Lambda }{2}}  \label{HamVsX}
\end{equation}%
where the last equality holds for the case $p_{i}|s_{pi,0}|=p_{i}|s_{pi,n}|=%
\sum p_{i}=0$. This is the centre-of-inertia coordinate system; we can
always choose it without loss of generality. Eqn. (\ref{HamVsX}) can always
be inverted to give X in terms of the Hamiltonian, 
\begin{equation}
X=\pm \frac{1}{\kappa }\sqrt{\frac{\kappa ^{2}H^{2}}{4}+\frac{\Lambda }{2}}.
\label{Ham}
\end{equation}%
Setting $X>0$\ by convention\footnote{%
Note that solutions with $X<0$ are equivalent to solutions with $X>0$ under
a parity transformation.} and substituting back into the recurrence
relations yields the determining equation for the Hamiltonian.

\bigskip

For $N=2$ the recurrence relations yield 
\begin{eqnarray}
&&\left( 4K_{1}^{2}+[\kappa \sqrt{p_{1}^{2}+m_{1}^{2}}-2K_{0}][\kappa \sqrt{%
p_{2}^{2}+m_{2}^{2}}-2K_{2}]\right) \mbox{tanh}\left( \frac{1}{2}%
K_{1}|z_{1}-z_{2}|\right)  \nonumber  \label{H2} \\
&&\makebox[5em]{}=-2K_{1}\left( [\kappa \sqrt{p_{1}^{2}+m_{1}^{2}}%
-2K_{0}]+[\kappa \sqrt{p_{2}^{2}+m_{2}^{2}}-2K_{2}]\right)
\end{eqnarray}%
which is the determining equation for the Hamiltonian as defined via eq. (%
\ref{Ham}). \ Note that in this particular case $K_{0}=K_{2}=\sqrt{\kappa
^{2}X^{2}-\frac{\Lambda }{2}}$ in the centre-of inertia system.

The $N=3$\ case is somewhat more tedious, though the procedure is
straightforward. \ The result is 
\begin{eqnarray}
&&\left[ \left( \left( \hat{M}_{1}+\hat{K}_{1}\right) \left( \hat{M}_{3}+%
\hat{K}_{4}\right) \hat{M}_{2}+\left( \hat{M}_{1}+\hat{K}_{1}\right) \hat{K}%
_{3}^{2}+\left( \hat{M}_{3}+\hat{K}_{4}\right) \hat{K}_{2}^{2}\right) \tanh
\left( \frac{\hat{K}_{3}}{4}z_{32}\right) \tanh \left( \frac{\hat{K}_{2}}{4}%
z_{21}\right) \right.   \nonumber \\
&&+\left( \left( \hat{M}_{1}+\hat{M}_{2}+\hat{K}_{1}\right) \left( \hat{M}%
_{3}+\hat{K}_{4}\right) +\hat{K}_{3}^{2}\right) \hat{K}_{2}\tanh \left( 
\frac{1}{4}\hat{K}_{3}z_{32}\right)   \nonumber \\
&&+\left( \left( \hat{M}_{1}+\hat{K}_{1}\right) \left( \hat{M}_{2}+\hat{M}%
_{3}+\hat{K}_{4}\right) +\hat{K}_{2}^{2}\right) \hat{K}_{3}\tanh \left( 
\frac{1}{4}\hat{K}_{2}z_{21}\right)   \nonumber \\
&&\left. +\left( \hat{M}_{1}+\hat{M}_{2}+\hat{M}_{3}+\hat{K}_{1}+\hat{K}%
_{4}\right) \hat{K}_{2}\hat{K}_{3}\right] =0  \label{det-eqn3}
\end{eqnarray}%
where%
\begin{eqnarray}
\hat{M}_{i} &=&\kappa \sqrt{p_{i}^{2}+m_{i}^{2}}  \label{new_def} \\
\hat{K}_{j} &=&-2\sqrt{\kappa ^{2}\left[ X+\frac{\epsilon }{4}\left(
\sum_{i=1}^{3}s_{pi,j}p_{i}\right) \right] ^{2}-\frac{\Lambda }{2}} 
\nonumber
\end{eqnarray}%
This system for the equal mass case is discussed in detail in ref. \cite%
{MikeSven}. When $\Lambda =0$\ \ this equation can be solved using a
generalization of the Lambert W-function \cite{MMS}.

\bigskip

When the $z_{i}$'s are not ordered such that $z_{n}<z_{n-1}<\ldots <z_{1}$,
as can occur when two particles cross one another, we can treat this
situation by relabelling the coordinates so that the ordering is in terms of
decreasing magnitude. The algorithm just described can then be applied. \ 

When the $z_{i}$'s are degenerate we are faced with a different situation. \
In quantum mechanics this corresponds to two charges merging into one. In
the case of two bodies (ie. H$_{2}^{+}$) there are two solutions, gerade
(symmetric) and ungerade (antisymmetric), whereas for the H atom (the single
Dirac delta function) there is only one solution. The gerade solution maps
onto the H atom solution continuously for both the energy and the
wavefunction. \ However the ungerade solution goes into the continuum near $%
R=1$, where $R$\ is the separation between the two nuclei in atomic units
(for more details see ref. \cite{ScottBabb}).\ It can be analytically
extended below $R=1$\ but it corresponds to a different branch of the
Lambert W-function. This latter energy does not map onto that for the
one-dimensional H atom. Furthermore the ungerade symmetry means the
wavefunction vanishes at the midpoint; as $R\rightarrow 0$\ this will not
reproduce the proper behaviour for the H atom. In one spatial dimension this
problem can be avoided by allowing the particles to cross as discussed in
refs. \cite{burnell} \ and \cite{Justin}.

\section{Discussion}

We have shown that the constraint equations of the relativistic
one-dimensional self-gravitating N--body problem can be solved exactly, and
have described the procedure for carrying this out. We have given explicit
results for the $N=2,3$ cases. \ We have also shown that this system is
formally equivalent to the stationary Schroedinger equation for a particle
of mass $m$ and energy $\frac{\hslash ^{2}}{2m}\mathcal{E}$ interacting with
a linear molecule whose field sources have charges $\kappa E_{pi}$. \ The
interpretation of the terms is different: in the gravitational case the
special-relativistic energy $\kappa E_{pi}$ is equivalent to the charge of
the field source.

For future work, the obvious problem to consider is quantization. Recent
progress on this for the two-body case in the post-Newtonian limit was
recently made, and the shift in energy levels caused by
quantum-gravitational effects was explicitly computed\cite{Matt}. In the
present case, since we already know the Lagrangian density whose
Euler-Lagrange equations are the Schroedinger wave equation, we can obtain a
Hamiltonian density and quantize the system. This procedure is akin to `2nd
quantization' ; it allows TRANSITIONS between states, with the dilaton
itself acting as the agent of transition.

Another avenue to explore is generalization of this work to higher
dimensions, ie to the 2+1 case or the 3+1 case. Since dimensional scaling
applies to H$_{2}^{+}$ \cite{Frantz,Lopez,ScottFrecon}, it is conceivable
that a scheme of dimensional scaling might apply to the gravity problem if
dilatons are involved. Though the generalization of the present work has not
yet been carried out for higher dimensions, we can nonetheless anticipate
the following features. The formulation herein yields in effect a
``decoupling'' of the system of integrodifferential equations of eqs.~(\ref%
{eqn-rt})-(\ref{eqn-stressenergy}) and solves for the dilation field, a
result not unlike the ``decoupling'' of Maxwell's equations leading to the
electromagnetic wave equation governing light. In view of the functional
forms of the action in eq.~(\ref{eqn-act1}) and the governing field equation
for the dilaton of eq.~(\ref{eqn-psi}), we can anticipate that the resulting
field equation will have at least a Laplacian term and terms corresponding
to an effective potential. These are the key terms for a Schr\"{o}dinger
equation. However, it is possible that extra terms, perhaps even non-linear
terms, may also appear. Non-linear terms could prevent a renormalizable
quantum theory but even a non-linear Schr\"{o}dinger equation - such as the
equation governing optical solitons - is manageable. At any rate, it is
quite also quite possible that the inclusion of an electromagnetic field, as
shown here for the $(1+1)$ case, may yield an equation linear insofar as the
electromagnetic terms are concerned and therefore a candidate for
quantization. At a minimum, a generalization to the $(2+1)$ case is bound to
be instructive.

The $(3+1)$ case introduces gravitons and consequently the physical
situation changes dramatically. However, this case is by no means hopeless.
It is possible through the use of domain walls that a $(3+1)$ dilatonic
formulation would also lead to a Schr\"{o}dinger equation. If so, the
(non-linear) effects of gravitational radiation could then be added on to a
quantized Schr\"{o}dinger equation, the situation ressembling the well-known
case of the hydrogen atom which collapses classically but remains stable
because the quantum principle prevents its energy from decreasing below the
ground state. At this stage no conclusions can be firmly drawn, though it
appears that a generalization to higher dimensions is certainly a worthwhile
investigation as it promises to be feasible to at least some degree.

\bigskip

{\huge Appendix\bigskip }

Here we \ derive the equations of motion for the N-body system. These can be
derived from Hamilton's equaions: 
\begin{eqnarray}
\dot{z}_{J} &=&\frac{\partial H}{\partial p_{J}}=\frac{\partial H}{\partial X%
}\frac{\partial X}{\partial p_{J}}=\frac{4\kappa X}{\sqrt{\kappa ^{2}X^{2}-%
\frac{\Lambda }{2}}}\frac{\partial X}{\partial p_{J}}  \label{EqnOfMotionZ}
\\
\dot{p}_{J} &=&-\frac{\partial H}{\partial z_{J}}=-\frac{\partial H}{%
\partial X}\frac{\partial X}{\partial z_{J}}=-\frac{4\kappa X}{\sqrt{\kappa
^{2}X^{2}-\frac{\Lambda }{2}}}\frac{\partial X}{\partial z_{J}}
\label{EqnOfMotionP}
\end{eqnarray}%
The only thing that needs to be done is to determine the partial derivatives
of $X$ with repect to $p_{a}$ and $z_{a}$. This can done by differentiating
eqn. (\ref{Ham}) in conjunction with the recurrence relations.

Before proceeding to the actual derivation, it is worthwhile cleaning up the
notation. Differentiating (\ref{KEqn}) with respect to $p_{a}$ and $z_{a}$
gives 
\begin{eqnarray}
\frac{\partial K_{r}}{\partial z_{J}} &=&\frac{\partial X}{\partial z_{J}}%
\frac{\kappa ^{2}(X+\frac{1}{4}\epsilon p_{i}s_{pi,r})}{K_{r}}=\frac{%
\partial X}{\partial z_{J}}D_{K_{r}} \\
\frac{\partial K_{r}}{\partial p_{J}} &=&\left( \frac{\partial X}{\partial
p_{J}}+\frac{1}{4}\epsilon s_{p_{J},r}\right) \frac{\kappa ^{2}(X+\frac{1}{4}%
\epsilon p_{i}s_{pi,r})}{K_{r}}=\left( \frac{\partial X}{\partial p_{J}}+%
\frac{1}{4}\epsilon s_{p_{J},r}\right) D_{K_{r}}
\end{eqnarray}%
Thus differentating the recurrence relations (\ref{RecRelationMA}) and (\ref%
{RecRelationMB}) with respect to $z_{j}$ gives

\begin{eqnarray}
\frac{\partial M_{A,n-r}}{\partial z_{J}} &=&2\frac{\partial X}{\partial
z_{J}}(D_{K_{n-r}}+D_{K_{n-r-1}})M_{A,n-r+1}e_{AA,n-r}{}  \nonumber \\
&&{}-2\frac{\partial X}{\partial z_{J}}%
(D_{K_{n-r}}-D_{K_{n-r-1}})M_{B,n-r+1}e_{AB,n-r}  \nonumber \\
&&{}+L_{AA,n-r}\frac{\partial M_{A,n-r+1}}{\partial z_{J}}e_{AA,n-r} 
\nonumber \\
&&{}+L_{AB,n-r}\frac{\partial M_{B,n-r+1}}{\partial z_{J}}e_{AB,n-r} 
\nonumber \\
&&{}+\frac{1}{2}\frac{\partial X}{\partial z_{J}}%
(D_{K_{n-r}}-D_{K_{n-r-1}})L_{AA,n-r}M_{A,n-r+1}e_{AA,n-r}  \nonumber \\
&&{}-\frac{1}{2}\frac{\partial X}{\partial z_{J}}%
(D_{K_{n-r}}+D_{K_{n-r-1}})L_{AB,n-r}M_{B,n-r+1}e_{AB,n-r}  \nonumber \\
&& \\
\frac{\partial M_{B,n-r}}{\partial z_{J}} &=&-2\frac{\partial X}{\partial
z_{J}}(D_{K_{n-r}}-D_{K_{n-r-1}})M_{A,n-r+1}e_{BA,n-r}{}  \nonumber \\
&&{}+2\frac{\partial X}{\partial z_{J}}%
(D_{K_{n-r}}+D_{K_{n-r-1}})M_{B,n-r+1}e_{BB,n-r}  \nonumber \\
&&{}-L_{BA,n-r}\frac{\partial M_{A,n-r+1}}{\partial z_{J}}e_{BA,n-r} 
\nonumber \\
&&{}-L_{BB,n-r}\frac{\partial M_{B,n-r+1}}{\partial z_{J}}e_{BB,n-r} 
\nonumber \\
&&{}-\frac{1}{2}\frac{\partial X}{\partial z_{J}}%
(D_{K_{n-r}}+D_{K_{n-r-1}})L_{BA,n-r}M_{A,n-r+1}e_{BA,n-r}  \nonumber \\
&&{}+\frac{1}{2}\frac{\partial X}{\partial z_{J}}%
(D_{K_{n-r}}-D_{K_{n-r-1}})L_{BB,n-r}M_{B,n-r+1}e_{BB,n-r},  \nonumber \\
&&
\end{eqnarray}%
Repeating for $p_{J}$ gives

\begin{eqnarray}
\frac{\partial M_{A,n-r}}{\partial p_{J}} &=&2\left( \frac{\partial X}{%
\partial p_{J}}+\frac{1}{4}\epsilon s_{p_{J},r}\right)
(D_{K_{n-r}}+D_{K_{n-r-1}})M_{A,n-r+1}e_{AA,n-r}{}  \nonumber \\
&&{}-2\left( \frac{\partial X}{\partial p_{J}}+\frac{1}{4}\epsilon
s_{p_{J},r}\right) (D_{K_{n-r}}-D_{K_{n-r-1}})M_{B,n-r+1}e_{AB,n-r} 
\nonumber \\
&&{}+L_{AA,n-r}\frac{\partial M_{A,n-r+1}}{\partial p_{J}}e_{AA,n-r} 
\nonumber \\
&&{}+L_{AB,n-r}\frac{\partial M_{B,n-r+1}}{\partial p_{J}}e_{AB,n-r} 
\nonumber \\
&&{}+\frac{1}{2}\left( \frac{\partial X}{\partial p_{J}}+\frac{1}{4}\epsilon
s_{p_{J},r}\right) (D_{K_{n-r}}-D_{K_{n-r-1}})L_{AA,n-r}M_{A,n-r+1}e_{AA,n-r}
\nonumber \\
&&-\frac{1}{2}\left( \frac{\partial X}{\partial p_{J}}+\frac{1}{4}\epsilon
s_{p_{J},r}\right) (D_{K_{n-r}}+D_{K_{n-r-1}})L_{AB,n-r}M_{B,n-r+1}e_{AB,n-r}
\nonumber \\
&& \\
\frac{\partial M_{B,n-r}}{\partial p_{J}} &=&-2\left( \frac{\partial X}{%
\partial p_{J}}+\frac{1}{4}\epsilon s_{p_{J},r}\right)
(D_{K_{n-r}}-D_{K_{n-r-1}})M_{A,n-r+1}e_{BA,n-r}{}{}  \nonumber \\
&&{}+2\left( \frac{\partial X}{\partial p_{J}}+\frac{1}{4}\epsilon
s_{p_{J},r}\right) (D_{K_{n-r}}+D_{K_{n-r-1}})M_{B,n-r+1}e_{BB,n-r} 
\nonumber \\
&&{}-L_{BA,n-r}\frac{\partial M_{A,n-r+1}}{\partial p_{J}}e_{BA,n-r} 
\nonumber \\
&&{}-L_{BB,n-r}\frac{\partial M_{B,n-r+1}}{\partial p_{J}}e_{BB,n-r} 
\nonumber \\
&&{}{}-\frac{1}{2}\left( \frac{\partial X}{\partial p_{J}}+\frac{1}{4}%
\epsilon s_{p_{J},r}\right)
(D_{K_{n-r}}+D_{K_{n-r-1}})L_{BA,n-r}M_{A,n-r+1}e_{BA,n-r}  \nonumber \\
&&{}+\frac{1}{2}\left( \frac{\partial X}{\partial p_{J}}+\frac{1}{4}\epsilon
s_{p_{J},r}\right)
(D_{K_{n-r}}-D_{K_{n-r-1}})L_{BB,n-r}M_{B,n-r+1}e_{BB,n-r},  \nonumber \\
&&
\end{eqnarray}

Combining these recurrence relations, the derivative of eqn. \ref{Ham} with
respect to $z_{J}$ and $p_{J}$, the equations 
\begin{equation}
\frac{\partial M_{B,1}}{\partial z_{J}}=0\qquad \text{and}\qquad \frac{%
\partial M_{B,1}}{\partial p_{J}}=0,
\end{equation}%
and the boundary conditions, 
\begin{equation}
M_{A,n+1}=M_{B,1}=0
\end{equation}%
will give a linear equation in $\frac{\partial X}{\partial z_{J}}$ and
another one in $\frac{\partial X}{\partial p_{J}}$ , which can always be
solved. Substituting back in eqns. (\ref{EqnOfMotionZ}) and (\ref%
{EqnOfMotionP}) will give Hamilton's equations in terms of $X$.

\pagebreak

{\Large Acknowledgements}

This work was supported in part by the Natural Sciences and Engineering
Research Council of Canada. P.S. Farrugia gratefully acknowledges support
from a Commonwealth Scholarship.

\end{document}